\documentclass[reprint]{revtex4-1}
\usepackage{amsmath,amssymb,graphics,graphicx,bbm,multirow,booktabs}

\usepackage{color,hyperref}
\hypersetup{colorlinks,breaklinks,
            linkcolor=blue,urlcolor=blue,
            anchorcolor=blue,citecolor=blue}

\newcommand{\hl}     {\ensuremath{T_{1/2}}}
\newcommand{\nubb}   {\ensuremath{0\nu\beta\beta}}
\newcommand{\nunubb}   {\ensuremath{2\nu\beta\beta}}

\newcommand{\mne} {\ensuremath{m_{\beta}}}
\newcommand{\mbb} {\ensuremath{m_{\beta\beta}}}
\newcommand{\senexp} {\ensuremath{\mathcal{E}}}
\newcommand{\senbkg} {\ensuremath{\mathcal{B}}}
\newcommand{\senroi} {\ensuremath{\text{ROI}}}

\newcommand{\medhl}  {\ensuremath{\hat{T}_{1/2}}} 
\newcommand{\medmbb} {\ensuremath{\hat{m}_{\beta\beta}}} 

\newcommand{\ctssen} {\ensuremath{\text{cts}/(\text{kg}_{iso}\,\senroi\,\text{yr})}}

\newcommand{\isot}[2]{\ensuremath{^{\text{#2}}}\text{#1}}

\begin{document}

\title{Discovery probability of next-generation neutrinoless double-$\beta$ decay experiments}

\author{Matteo Agostini}
\email{matteo.agostini@gssi.infn.it}
\affiliation{Gran Sasso Science Institute, L'Aquila, Italy}
\author{Giovanni Benato}
\email{gbenato@berkeley.edu}
\affiliation{Department of Physics, University of California, Berkeley, CA 94720 - USA\\
  Nuclear Science Division, Lawrence Berkeley National Laboratory, Berkeley, CA 74720 - USA}
\author{Jason A. Detwiler}
\email{jasondet@uw.edu}
\affiliation{Center for Experimental Nuclear Physics and Astrophysics, and Department of Physics, University of Washington, Seattle, WA 98115 - USA}
\author{~}

\begin{abstract}
\noindent
The Bayesian discovery probability of future experiments searching for
neutrinoless double-$\beta$ decay is evaluated under the popular assumption that
neutrinos are their own antiparticles.
A Bayesian global fit is performed to construct a probability
distribution for the effective Majorana mass, the observable of interest for these
experiments. 
This probability distribution is then combined with the sensitivity of each
experiment derived from a heuristic counting analysis.
The discovery probability is found to be higher than previously considered, but
strongly depends on whether the neutrino mass ordering is normal or inverted.
For the inverted ordering, next-generation experiments are likely
to observe a signal already during their first operational stages.
Even for the normal ordering, in the absence of neutrino mass mechanisms
that drive the lightest state or the effective Majorana mass to zero,
the probability of discovering neutrinoless double-$\beta$ decay can reach
$\sim$50\% or more in the most promising experiments.
\end{abstract}

\date{\today}
\maketitle

\section{Introduction}
Definitive evidence for non-zero neutrino masses from oscillation experiments
has been available for nearly two
decades~\cite{Kajita:2016cak,McDonald:2016ixn,Eguchi:2002dm,Olive:2016xmw}. 
However, the incorporation of neutrino masses into the Standard Model (SM) of 
particle physics remains an open issue.
Because it is electrically neutral, the neutrino is the only known fundamental fermion
that could be its own anti-particle, 
and obtain its mass through a Majorana mass term~\cite{Majorana2006}.
Such a Majorana mass term would
violate total lepton number conservation, and naturally emerges in many
beyond-the-SM theories~\cite{Vergados:2012xy}. It also emerges in leading theories
that explain the dominance of matter over antimatter in the universe~\cite{Fukugita:1986hr},
to which we owe our very existence. The motivation to test the Majorana nature 
of the neutrino has never been higher.

At present, the only feasible method for testing a pure-Majorana SM neutrino
without requiring new fields or symmetries is to search for neutrinoless
double-$\beta$ (\nubb) decay~\cite{Kayser:2005cy}. In this hypothetical nuclear transition
a nucleus of mass number $A$ and charge $Z$ decays
as $(A, Z)\rightarrow(A, Z+ 2)+ 2e^-$~\cite{Furry:1939qr}.
A positive detection would signify the first observation of a matter-creating process
(without the balancing emission of antimatter),
and would unambiguously establish that neutrinos have
a Majorana mass component, independent of the channels involved in the transition or
the isotope under study~\cite{Schechter:1981bd,Duerr:2011zd}.
An experimental campaign to search for this process has been underway for
decades, and its continuation requires
intense effort and significant resources. 
We set out to explore the justification for such an expenditure,
a task for which Bayesian methods are particularly well suited.
In this work we present our evaluation, using all available information about neutrino
phenomenology, of the Bayesian probability that future
\nubb\ decay searches will prove that neutrinos are
Majorana particles.

Neutrino phenomenology is described by an extension
of the SM in which three
quantum flavor states $\nu_e$, $\nu_{\mu}$, and
$\nu_{\tau}$ couple to charged leptons via the weak interaction~\cite{Olive:2016xmw}. 
Such flavor states do not have a fixed mass but are rather a
quantum-mechanical superposition of three mass eigenstates $\nu_1$,
$\nu_2$, $\nu_3$, with masses $m_1$, $m_2$, $m_3$.
The transformation between the mass and flavor bases is described
by the unitary PMNS matrix,
which is parametrized by three mixing angles ($\theta_{12}$, $\theta_{13}$, $\theta_{23}$),
the CP-violating phase $\delta$, and two Majorana phases ($\alpha_{21}$, $\alpha_{31}$).
Consequently, neutrinos can transform from one flavor state to another
during propagation, giving rise to neutrino oscillation, which remains to date 
the only observed phenomenon requiring non-zero neutrino masses.
The transformation probability is a function of the
two squared mass differences $\Delta m_{31}^2$ and $\Delta m_{21}^2$ 
(where $\Delta m_{ij}^2 \equiv m_i^2 - m_j^2$), the three mixing angles, and $\delta$.
All oscillation parameters have been measured with the exception of
$\delta$ and the sign of $\Delta m_{31}^2$~\cite{Olive:2016xmw,Esteban:2016qun}.
These parameters should be accessible in the near future 
by experiments exploiting vacuum oscillations or matter-induced 
flavor transformations~\cite{Adamson:2017gxd,Abe:2017uxa,Blennow:2013oma,Capozzi:2017ipn}.

Complementary constraints on neutrino phenomenology are provided by
cosmological observations, precision measurements of $\beta$-decay kinematics, and
\nubb\ decay searches~\cite{Rodejohann:2011mu,Zhang:2015kaa}.
Lepton number violation searches at accelerators 
and other experimental probes can provide additional information on neutrinos 
(see, for example, \cite{Deppisch:2015qwa,Peng:2015haa}) but will not be discussed here.

Cosmology is sensitive to the sum of neutrino masses:
$\Sigma=m_1+m_2+m_3$.
The value of $\Sigma$ is constrained by 
Planck and other observations~\cite{Ade:2015xua} which set upper limits on the order of tens
to hundreds of meV, depending on the model and data-sets used for
analysis~\cite{DellOro:2015kys}. A lower limit for $\Sigma$ is imposed
by the measurements of the mass splittings.

The energy spectrum end-point of electrons emitted in nuclear
$\beta$-decay is sensitive to the rest mass of the electron antineutrino~\cite{Fermi:1934hr,Fermi:1934sk}. 
Since the neutrino mass splittings are smaller than can be resolved with available electron spectroscopic
techniques, the end point defect is characterized by the effective neutrino mass:
\begin{equation}
   \mne \equiv \sqrt{m_1^2\,c_{12}^2\,c_{13}^2 
                              + m_2^2\,s_{12}^2\,c_{13}^2
                              + m_3^2\,s_{13}^2}
\end{equation}
where $s_{ij} = \sin \theta_{ij}$ and $c_{ij} = \cos \theta_{ij}$.
A non-zero \mne\ has not yet been observed and the best upper limits are set by the Troitsk~\cite{Aseev:2011dq}
and Mainz~\cite{Kraus:2004zw} experiments, giving $\mne<2.12$\,eV at 95\% credible interval (CI)
and $\mne<2.3$\,eV at 95\% confidence level (CL), respectively.

The strongest limits on the half life of \nubb\ decay are from the
KamLAND-Zen~\cite{KamLAND-Zen:2016pfg} and GERDA~\cite{Agostini:2017iyd} experiments, giving
$\hl(\isot{Xe}{136})>10.7 \times 10^{25}$\,yr (sensitivity: $5.6\times10^{25}$\,yr)
and $\hl(\isot{Ge}{76})>5.3 \times 10^{25}$\,yr (sensitivity: $4.0\times10^{25}$\,yr)
at 90\% CL, respectively.
In a minimal SM extension that incorporates neutrino masses
by only adding Majorana neutrino mass terms
for the three known mass eigenstates to the SM Lagrangian, 
\nubb\ decay is mediated by the exchange of neutrinos.
In this case, the half life of the process is given by~\cite{Engel:2016xgb}:
\begin{equation}
   (\hl)^{-1}  = G_{0\nu} \, |\mathcal{M}_{0\nu}|^2 \, \mbb^2
   \label{eq:t12tombb}
\end{equation}
where $G_{0\nu}$ is a phase-space factor and $\mathcal{M}_{0\nu}$ is the
dimensionless nuclear matrix element (NME) encompassing the nuclear
physics. The observable \mbb, the effective Majorana mass, is given by
\begin{equation}
   \mbb
   \equiv 
   \left| 
        m_1\,c_{12}^2\,c_{13}^2
      + m_2\,s_{12}^2\,c_{13}^2\,e^{i\alpha_{21}}
      + m_3\,s_{13}^2 \,e^{i(\alpha_{31}-\delta)}
   \right|
   \label{eq:mbb}
\end{equation}
The aforementioned limits on \hl\ translate
to  $\mbb<61-165$\,eV and $\mbb<150-330$\,eV (90\%
CL)~\cite{KamLAND-Zen:2016pfg,Agostini:2017iyd}.
The ranges account for different theoretical calculations of the 
NME~\cite{Engel:2016xgb}.

For light Majorana neutrino exchange, the allowed parameter space for \nubb\ decay is considerably constrained. 
In the case $\Delta m_{31}^2 < 0$, referred to as the inverted neutrino mass ordering (IO),
the oscillation parameters dictate that \mbb\ cannot be much lower than $\sim$18\,meV~\cite{Capozzi:2017ipn}. 
A broad international experimental program requiring considerable resources is being 
mounted to search for \nubb\ decay in this range. These experiments will also be sensitive
to part of the parameter space for the normal ordering (NO, corresponding to $\Delta m_{31}^2 > 0$),
although if \mbb\ is exceedingly small even larger experiments will be required.

To maximize the return on this investment,
it is the opinion of the authors that the design of these future experiments should be driven by 
the likelihood of discovering \nubb\ decay, rather than limit-setting capability as is usually done.
With the aim of furthering progress in this direction,
this article presents a global Bayesian analysis to extract the present-day
probability distribution of \mbb\ using all relevant experimental information available to date.
The probability distribution is folded with the discovery sensitivity of
future \nubb\ decay experiments, computed here with a heuristic counting analysis.
As will be seen, the resulting discovery probabilities indicate that next-generation
experiments have a high likelihood of observing a signal if neutrinos are indeed Majorana particles.

Our analysis focuses on scenarios in which the lightest neutrino mass
eigenstate $m_l$ and \mbb\ are not fixed by mass mechanisms or
flavour symmetries that would significantly alter the parameter space of
interest~\cite{Feruglio:2002af,King:2013psa,Agostini:2015dna}.
To explore the discovery probability of models yielding hierarchical mass
spectra (i.e. with $m_l \ll m_2$~\cite{Nath:2016mts}), 
we analyze the extreme
case in which $m_l$ is zero. 
Projections for models predicting intermediate values of $m_l$ can be obtained
by interpolating between our results.
We do not consider explicitly scenarios in which \mbb\ has a fixed value since the
posterior distribution would be a trivial delta function and the discovery
probability can be directly extracted from the sensitivity of the experiments.

\section{Global fit}
\label{sec:globfit}
The parameter basis selected for our global fit is
$\{\Sigma,$ 
$\Delta m_{21}^2,$
$\Delta m_{31}^2$ or $\Delta m_{23}^2,$
$\theta_{12},$ 
$\theta_{13},$
$\alpha_{21},$
$(\alpha_{31}-\delta)\}$,
where
$\Delta m_{31}^2$ is used for NO and $\Delta m_{23}^2$ for IO.
The notation is taken from Ref~\cite{Olive:2016xmw}.
The remaining degrees of freedom of the model do not affect the analysis and are
neglected. Statistical correlations of the parameters in the basis are also
negligible~\cite{Esteban:2016qun}.
The ignorance on the scale of the parameters is introduced through scale
invariant priors: the priors of the mass observables are logarithmic
whereas the priors of angles and phases -- whose values are restricted to the range
$[0,2\pi]$ --  are flat.

The choice of the basis affects our results only slightly as long as the basis
covers all degrees of freedom of the problem and its parameters are constrained
by the data (see discussion in \appendixname~\ref{sec:fitDetails}).
Our usage of $\Sigma$ and the mass splittings
to cover the three degrees of freedom related to the neutrino masses
is motivated both by physical and statistical arguments.
The $\Delta m^2$ parameters are direct
observables of oscillation experiments and $\Sigma$ can be physically interpreted
as the neutrino mass scale.
In addition, $\Sigma$ is constrained by the data to a finite range and
cannot vanish. This is critical for having a normalizable posterior distribution
when a scale-invariant prior is employed.

This basis does not accommodate scenarios with extreme hierarchical mass spectra
in which $m_l$ is driven to zero. 
The discovery probability for such hierarchical scenarios can however be easily studied by
fixing $\Sigma$ to its lower limit.
It might seem appropriate to to use a basis in which 
$\Sigma$ is substituted by $m_l$. This apparently natural choice 
poses serious difficulties, as the data are not directly sensitive to
$m_l$ except in the case of quasi-degenerate masses.
If used as an element of the basis, its posterior would be strongly
influenced by whatever prior is chosen in the low-mass range. 
The use of scale-invariant priors would then make
the \mbb\ probability distribution non-normalizable without the imposition of an
ad-hoc cutoff on $m_l$, whose value affects directly the posteriors.
The choice of a cutoff could be motivated by theoretical reasons (see
e.g. Ref.~\cite{Davidson:2006tg}), but it would still insert into the analysis an
arbitrary assumption which affects the results of the fit.

The likelihood function of the available data is constructed as the product of normalized factors, each
expressing the conditional probability of a sub-set of data given the value of
an observable:
\begin{equation}
\begin{array}{lrl}
\mathcal{L} & = & \mathcal{L}(\mathcal{D}_{\text{osc}}| \Delta m_{21}^2)
 \cdot  \mathcal{L}(\mathcal{D}_{\text{osc}}| \Delta m_{31}^2/\Delta m_{23}^2)\\
&\cdot& \mathcal{L}(\mathcal{D}_{\text{osc}}| s_{12}^2)
 \cdot  \mathcal{L}(\mathcal{D}_{\text{osc}}| s_{13}^2)\\
&\cdot& \mathcal{L}(\mathcal{D}_{\text{Troitsk}}|\mne)
 \cdot  \mathcal{L}(\mathcal{D}_{\text{\nubb}}|\mbb).
\end{array}
\end{equation}
where:
$\mathcal{D}_{\text{osc}}$ are the oscillation
data, whose likelihoods are computed using the nu-fit analysis (v3.0, Nov.~2016)~\cite{Esteban:2016qun};
$\mathcal{D}_{\text{Troitsk}}$ refers to the limit from
Troitsk~\cite{Aseev:2011dq} (including also the limit from Mainz~\cite{Kraus:2004zw} 
yields no perceptible change in the \mbb\ distribution);
and $\mathcal{D}_{\text{\nubb}}$ is the combined results from GERDA and
KamLAND-Zen. The latter is built using the sensitivity of the experiments rather
then their actual limits, which are strengthened by background under-fluctuations (that is, we
consider power-constrained limits~\cite{Cowan:2011an}). This results in a
normalized exponentially-decreasing likelihood with 90\% quantile at
$\mbb =  71-161$\,meV, depending on the choice of NME.

The NME values are fixed parameters in this analysis and the impact of their
variation is evaluated by performing the calculations multiple times assuming
different nuclear models. We consider the quasi-particle 
random phase approximation (QRPA~\cite{Simkovic:2013qiy,Hyvarinen:2015bda,Mustonen:2013zu}),
the interacting shell model (ISM~\cite{Menendez:2008jp,Horoi:2015tkc}),
the interacting boson model (IBM-2~\cite{Barea:2015kwa}), and
energy density functional theory (EDF~\cite{Vaquero:2014dna,Yao:2014uta}).
Within each model we use the average of the computations performed by different groups,
taking the spread as an indication of systematic uncertainty as discussed in
Ref.~\cite{Engel:2016xgb}.
We perform the primary analysis without considering quenching of the axial vector coupling constant $g_A$.
The effect of variation of $g_A$ is discussed below.
For recent insight into the status of the quenching issue we refer the reader to Ref.~\cite{Engel:2016xgb}.

\begin{figure*}[tb]
   \centering
   \includegraphics[width=\textwidth]{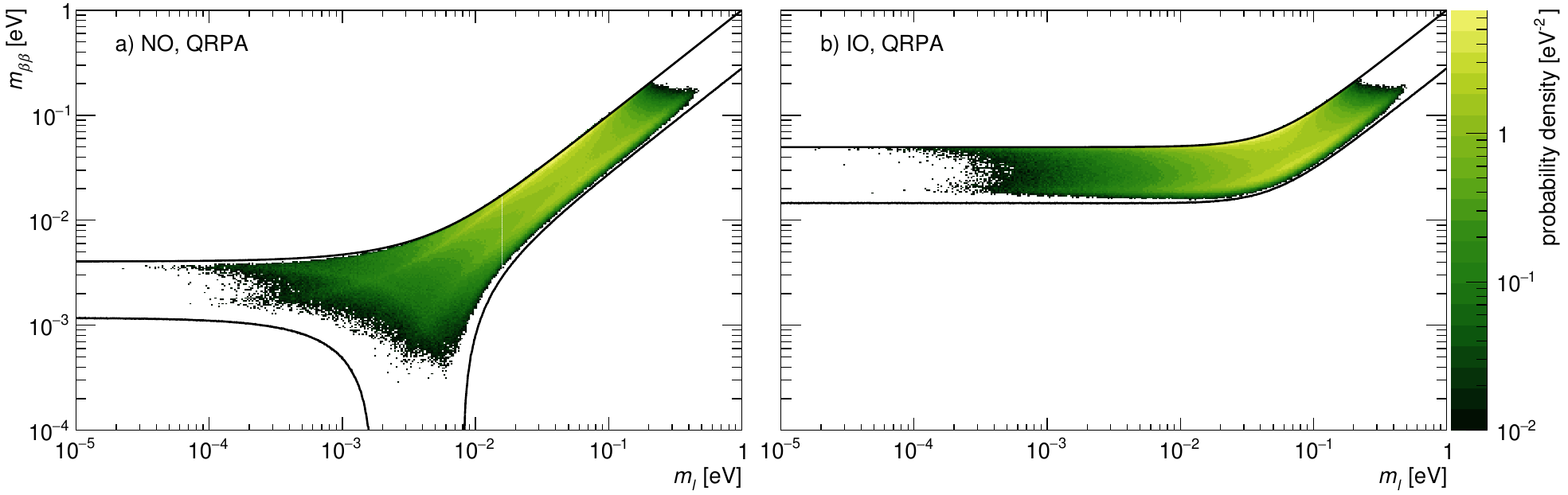}
   \caption{Marginalized posterior distributions for \mbb\ and $m_l$
      for NO (a) and IO (b). The solid lines show the allowed
      parameter space assuming 3$\sigma$ intervals of the
      neutrino oscillation observables from nu-fit~\cite{Esteban:2016qun}.  The
      plot is produced assuming QRPA NMEs and the absence of mechanisms
      that drive $m_l$ or \mbb\ to zero.
      The probability density is normalized by the logarithm of \mbb\ and of $m_l$.}
   \label{fig:mbb_vs_mlight}
\end{figure*}

The marginalized posterior distributions for all parameters of the basis and
observables of interest are computed via Markov-chain Monte-Carlo numerical
integrations with the BAT toolkit~\cite{Caldwell20092197}. All marginalized
distributions are included in \appendixname~\ref{sec:fitDetails}.
The posterior distributions for \mbb\ as a function of $m_l$
are shown in \figurename~\ref{fig:mbb_vs_mlight}, 
separately for the NO and IO scenarios.
The color map indicates the probability density and 
the solid lines show the maximally allowed parameter space given the
constraints on the oscillation parameters from nu-fit.
The volume of the allowed parameter space is dominated by the freedom of the
Majorana phase values, on which no direct measurement is available. 
The probability density is clearly nonuniform:
high \mbb\ values are disfavored by
the experimental limits on \mbb\ and \mne;
low \mbb\ values are unlikely because a fine tuning of the Majorana phases is
needed for the right-hand-side of equation~\eqref{eq:mbb} to vanish~\cite{Benato:2015via}. 
$m_l$ is unlikely to assume low values 
because oscillation experiments constrain $\Sigma$ to be larger than $|\Delta m_{31}^2|$,
and its scale invariant prior leaves a small volume of probable parameter space near that lower bound
over which $m_l$ can become small.
Our results are consistent with previous
work~\cite{Benato:2015via,Gerbino:2015ixa,Gerbino:2016ehw} and the differences can be
attributed to the different data sets considered.

\begin{figure}[tb]
   \centering
   \includegraphics[width=\columnwidth]{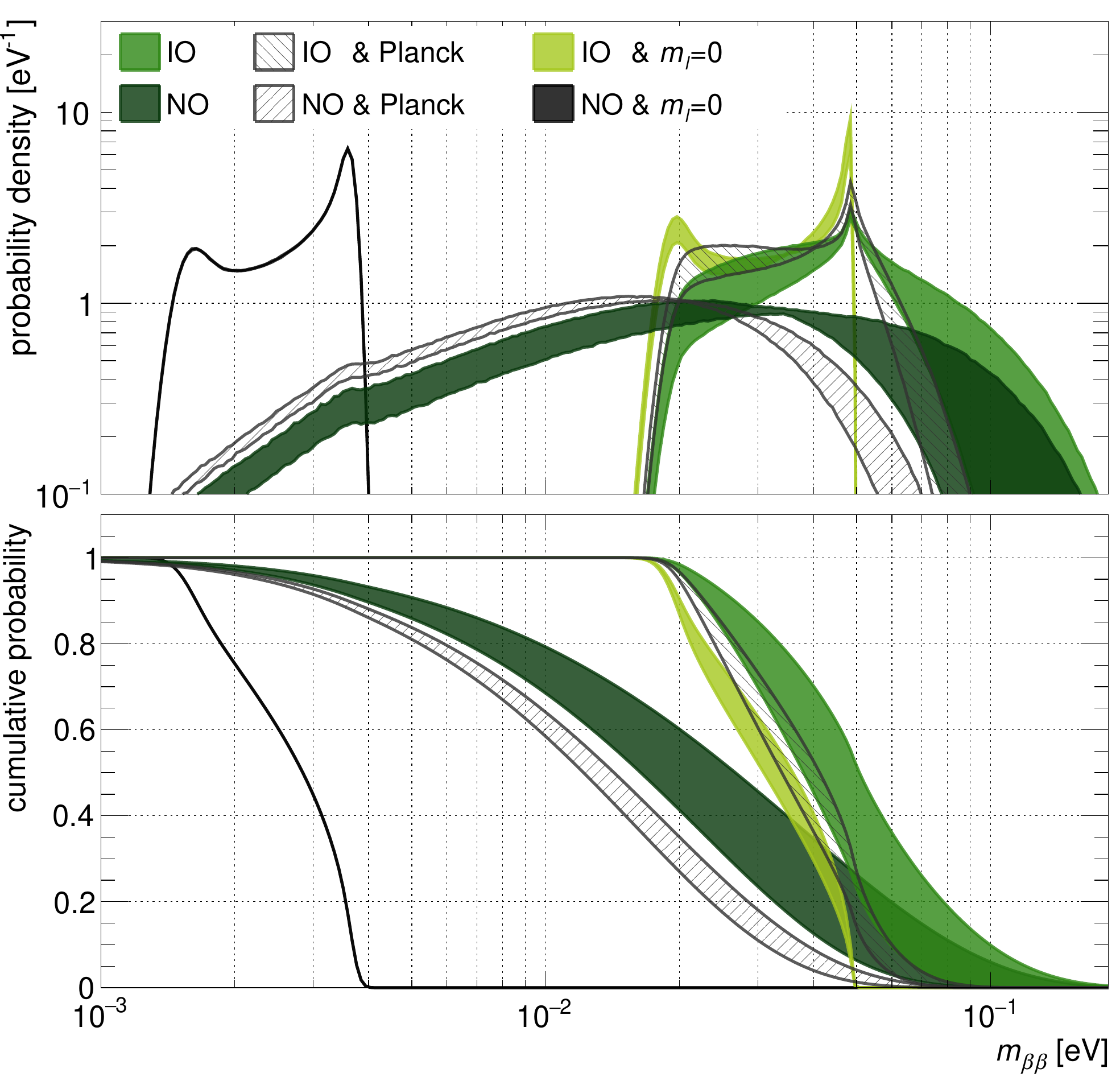}
   \caption{Top: marginalized posterior distributions of \mbb\ (solid line)
      for NO and IO, normalized by the logarithm of \mbb.
      Bottom: complementary cumulative distribution functions for \mbb.
      The band shows the deformation of the posterior distribution
      due to different assumptions on the NME.
      The data from cosmology provide a somewhat stronger constraint on \mbb\ than the current
      \nubb\ decay experiments. 
      The sharp peaks visible in the \mbb\ distributions 
      are due to a volume effect dominated by $\Sigma$ and the Majorana phases.
      For NO with $m_l = 0$ there is negligible variation due to the NME.
    }
   \label{fig:posteriors}
   %
\end{figure}

\figurename~\ref{fig:posteriors} shows the marginalized posterior distributions
for \mbb\ and the corresponding cumulative distributions.
The deformation of the posterior distributions due to the NME is visualized by
the band.
%
%
The 90\% probability central interval for \mbb\ is 20-119\,meV assuming IO and
3-104\,meV assuming NO, where we have allowed for the maximum variation
among the various NME considered. 
Consequently, the next-generation experiments that aim
for a discovery sensitivity of $10-20$\,meV will cover the true value of \mbb\ with
$>95\%$ probability assuming IO and with $\sim$50\% probability assuming NO.
To cover the true value of \mbb\  in the case of NO with about 90\% probability, an
experiment should reach a discovery sensitivity of about 5\,meV.

\figurename~\ref{fig:posteriors} shows also the distributions 
constructed when the data from Planck and other cosmological observations are added to the fit as an additional
normalized factor of the likelihood
$\mathcal{L}(\mathcal{D}_{\text{cosm}}|\Sigma)$, where
$\mathcal{D}_{\text{cosm}}$ represents the observational constraints on $\Sigma$ from the
combination of data labeled as ``TT+lowP+lensing+ext'' in Ref.~\cite{Ade:2015xua}.
These new data disfavor the quasi-degenerate region at high values of $m_l$
and compress the distributions of \mbb\ to lower values.
In this work, we use as reference results those obtained without imposing cosmological constraints.
This choice is motivated by the fact that cosmological constraints
are model-dependent, not only on the $\Lambda$CDM model used to interpret the data, but also on a host
of astrophysical models required to extract limits on $\Sigma$ from disparate datasets with
complex and interrelated systematic uncertainties~\cite{DellOro:2015kys,Palanque-Delabrouille:2014jca}.
In any case, at present the impact of cosmological data is still limited: the cumulative distributions
of \mbb\ -- and ultimately also the experimental discovery probabilities  --  change
by only tens of percent. 

When the fit is performed with $\Sigma$ fixed to its minimum allowed value
(corresponding to $m_l=0$),
the \mbb\ posterior distribution is constrained to lie within the horizontal bands
that extend to $m_l \rightarrow 0$ in \figurename~\ref{fig:mbb_vs_mlight}.
The \mbb\ posterior distribution is slightly shifted to smaller values for IO,
and the discovery probability of future experiments remains very high.
In NO, \mbb\ is pushed below the reach of future experiments, and the discovery 
probabilities are driven to be very small as shown in
\figurename~\ref{fig:posteriors}.
Using $m_l$ in the fit basis with a log-flat scale invariant prior would
provide the same results as long as the cutoff on $m_l$, required to have
normalizable posterior distributions, is set low enough to make the
result independent of the choice of cutoff.

\section{Experimental sensitivity}
The experimental search for \nubb\ decay is a very active field.  There is
a number of isotopes that can undergo \nubb\ decay and many detection 
techniques have been developed and tested in recent years~\cite{Cremonesi:2013vla,DellOro:2016tmg}. 
Examples are: high-purity Ge detectors~\cite{Ackermann:2012xja,Abgrall:2013rze},
cryogenic bolometers~\cite{CUORE-NIM2004,Artusa:2014lgv},
loaded organic liquid scintillators~\cite{KamLAND-Zen:2016pfg},
time-projection chambers~\cite{Albert:2014awa,ferrario}, and
tracking chambers~\cite{Arnold:2004xq}.
Various larger-scale experiments with the sensitivity to probe the full
IO parameter space are being mounted or proposed for the near or far future.
This work focuses on those projects considered recently by 
the U.S. DOE/NSF
Nuclear Science Advisory Committee's Subcommittee on Neutrinoless
Double Beta Decay~\cite{NLDBD}:
CUPID~\cite{Wang:2015raa,Wang:2015taa}, 
KamLAND-Zen~\cite{shirai},
LEGEND~\cite{bernhard,stefan},
nEXO~\cite{Mong:2016sza},
NEXT~\cite{Alvarez:2012flf},
PandaX-III~\cite{Chen:2016qcd},
SNO+~\cite{Lozza:2016rwo,KleinSNO2}, and 
SuperNEMO~\cite{Povinec:2017trz,Arnold:2010tu}.
Most of these projects follow a staged-approach in which the target mass will be
progressively increased. The various phases and parameters of each project are
summarized in \tablename~\ref{tab:par} and discussed in
\appendixname~\ref{sec:parameters}.
We would like to caution the reader, however, that 
many of these experiments are under rapid development, and the
parameters publicly available during the snapshot of time in which this manuscript
was prepared will often poorly characterize their ultimate reach.
Our conclusions should therefore be taken with a heavy grain of salt,
and we implore the reader to resist the urge to use our results to
make comparisons between experiments, and instead to focus on their
combined promise as a global, multi-isotope endeavor. We hope that our methods
are also useful as a figure-of-merit by which individual experiments can
evaluate their own implementations. 
This analysis will be updated when new information becomes available.

A primary experimental signature for \nubb\ decay  is
a mono-energetic peak in the measured energy spectrum at the $Q$-value of the decay,
produced when the two electrons emitted in the process are fully absorbed in the
detector's active volume.
While in many detectors additional analysis handles are available to distinguish
signal from background, energy is the one observable that is both necessary and sufficient
for discovery, and so the sensitivity of a \nubb\ decay experiment is
driven by Poisson statistics for events near the $Q$-value. It can thus be approximated with a
heuristic counting analysis, where there are just two parameters of interest: the
``sensitive exposure'' (\senexp) and the ``sensitive background''
(\senbkg). 
\senexp\ is given by the product of active isotope mass 
and live time, corrected by the active fiducial volume, the signal detection
efficiency, and the probability for a \nubb\ decay event to fall
in the energy region of interest (\senroi) in which the experiment is sensitive to the
signal. \senbkg\ is the number of background events in the \senroi\ after all analysis
cuts divided by \senexp.
The number of signal and background counts in the final spectrum
is then given by:
\begin{equation}
   N_{\nubb} = \dfrac{\ln 2 \cdot N_{A} \cdot \senexp}{ m_a \cdot\hl} \qquad
   \text{and} \qquad N_{bkg}= \senbkg\cdot \senexp
\end{equation}
where $N_A$ is Avogadro's number,  $m_a$ is the
molar mass of the target isotope, and \hl\ is the half-life of the decay.

The experimental efficiencies can be separated into: 
the actual fraction of mass used for analysis $\epsilon_{FV}$
(accounting for dead volumes in solid detectors and fiducial volume cuts
in liquid and gaseous detectors), 
the signal efficiency $\epsilon_{sig}$ (which is the product of the analysis
cut efficiency and the $\nubb$ containment efficiency), and 
the fraction of fully-contained $\nubb$ decay events with energy reconstructed in the \senroi. 
The choice of optimal \senroi\ depends on the background rate, its energy distribution, and
the energy resolution ($\sigma$) of the Gaussian peak expected from the signal.
Experiments with an excellent energy resolution ($\sigma<1\%$) have a \senroi\
centered at the $Q$-value with a width depending on the background rate.
For experiments with poorer energy resolution, the background
due to two-neutrino double-$\beta$ decay is significant up to the
$Q$-value. These experiments have an asymmetric optimal \senroi\ covering
primarily the upper half of the Gaussian signal. Our method to compute the
optimal \senroi\ is discussed in \appendixname~\ref{sec:heuristics}.

%
\begin{table*}[]
   \centering
   \caption{Experimental parameters of next-generation experiments.  The quoted
      mass refers to the \nubb\ decaying isotope and the energy resolution to
      the standard deviation ($\sigma$).  The \senroi\ edges are given in units
      of $\sigma$ from the Q-value of the decay.  $\epsilon_{FV}$ is the
      fraction of mass used for analysis and $\epsilon_{sig}$ is the signal
      detection efficiency. For SuperNEMO only, the reported $\epsilon_{sig}$ encompasses also
      the fraction of \nubb\ decay events in the \senroi.
      The sensitive exposure (\senexp) and background
      (\senbkg) are normalized to 1\,yr of live time.  \medhl\  and \medmbb\ are the
      median 3$\sigma$ discovery sensitivities assuming 5\,years of live time.  The
      \medmbb\  ranges account for the different NME calculations considered in the
      analysis.  The last columns show the envisioned reduction of background level and
      $\sigma$, as well as the expected increase of isotope mass, with respect to predecessor experiments
      which have released data at the time of manuscript preparation; ``n/a''
      indicates that no published experimental data are available yet.}
   \begin{tabular}{l|c|c|c|c|c|c|c|c|c|c|c|c|c}
     \hline
     \multirow{2}{*}{Experiment} & \multirow{2}{*}{Iso.} & Iso. &
     \multirow{2}{*}{$\sigma$} & \multirow{2}{*}{\senroi}  & \multirow{2}{*}{$\epsilon_{FV}$} &
     \multirow{2}{*}{$\epsilon_{sig}$} & \multirow{2}{*}{\senexp} & \multirow{2}{*}{\senbkg} &
     \multicolumn{2}{c|}{$3\sigma$ disc. sens.} & \multicolumn{3}{c}{Required} \\
     & & Mass & & & & & & & \medhl & \medmbb & \multicolumn{3}{c}{Improvement} \\
                &      & \multirow{2}{*}{[kg$_{iso}$]} & \multirow{2}{*}{[keV]}    & \multirow{2}{*}{[$\sigma$]} & \multirow{2}{*}{[$\%$]}
                & \multirow{2}{*}{[$\%$]}           
                & \multirow{2}{*}{$\left[ \dfrac {\text{kg}_{iso}\,\text{yr}} {\text{yr}} \right]$}
                & \multirow{2}{*}{$\left[ \dfrac {\text{cts}}{\text{kg}_{iso}\,\senroi\,\text{yr}} \right]  $}
     & \multirow{2}{*}{[yr]}   & \multirow{2}{*}{[meV]}    & \multirow{2}{*}{Bkg} & \multirow{2}{*}{$\sigma$} & Iso. \\
     & & & & & & & & & & & & & Mass\\
     \hline
     LEGEND 200~\cite{bernhard,stefan}               & $^{\ 76}$Ge & $175$   & $1.3$  & $[\mbox{-}2,2]$     & $93$  & $77$             & $119$             & $1.7\cdot10^{\mbox{-}3}$ & $8.4\cdot10^{26}$ & $40$--$73$  & $3$ & $1$ & 5.7 \\
     LEGEND 1k~\cite{bernhard,stefan}                & $^{\ 76}$Ge & $873$   & $1.3$  & $[\mbox{-}2,2]$     & $93$  & $77$             & $593$             & $2.8\cdot10^{\mbox{-}4}$ & $4.5\cdot10^{27}$ & $17$--$31$  & $18$  & $1$ & 29 \\
     SuperNEMO~\cite{Povinec:2017trz,Arnold:2010tu}  & $^{\ 82}$Se   & $100$   & $51$   & $[\mbox{-}4,2]$            & $100$ & $16$             & $16.5$            & $4.9\cdot10^{\mbox{-}2}$ & $6.1\cdot10^{25}$ & $82$--$138$ & $49$  & $2$ & 14 \\
     CUPID~\cite{Artusa:2014wnl,Wang:2015raa,Wang:2015taa}        & $^{\ 82}$Se & $336$   & $2.1$  & $[\mbox{-}2,2]$     & $100$ & $69$             & $221$             & $5.2\cdot10^{\mbox{-}4}$ & $1.8\cdot10^{27}$ & $15$--$25$  & n/a    & $6$ & n/a \\
     CUORE~\cite{CUORE-NIM2004,Artusa:2014lgv}        & $^{130}$Te & $206$   & $2.1$  & $[\mbox{-}1.4,1.4]$     & $100$ & $81$             & $141$             & $3.1\cdot10^{\mbox{-}1}$ & $5.4\cdot10^{25}$ & $66$--$164$  & $6$    & $1$ & 19 \\
     CUPID~\cite{Artusa:2014wnl,Wang:2015raa,Wang:2015taa}        & $^{130}$Te  & $543$   & $2.1$  & $[\mbox{-}2,2]$     & $100$ & $81$             & $422$             & $3.0\cdot10^{\mbox{-}4}$ & $2.1\cdot10^{27}$ & $11$--$26$  & $3000$ & $1$ & 50 \\
     SNO$+$ Phase\,I~\cite{Lozza:2016rwo,singh}               & $^{130}$Te  & $1357$  & $82$   & $[\mbox{-}0.5,1.5]$ & $20$  & $97$             & $164$             & $8.2\cdot10^{\mbox{-}2}$ & $1.1\cdot10^{26}$ & $46$--$115$ & n/a     & n/a & n/a \\
     SNO$+$ Phase\,II~\cite{KleinSNO2}               & $^{130}$Te  & $7960$  & $57$   & $[\mbox{-}0.5,1.5]$ & $28$  & $97$             & $1326$             & $3.6\cdot10^{\mbox{-}2}$ & $4.8\cdot10^{26}$ & $22$--$54$ & n/a     & n/a & n/a \\
     KamLAND-Zen 800~\cite{shirai}                            & $^{136}$Xe  & $750$   & $114$  & $[0,1.4]$   & $64$  & $97$             & $194$             & $3.9\cdot10^{\mbox{-}2}$ & $1.6\cdot10^{26}$ & $47$--$108$ & $1.5$  & $1$ & 2.1 \\
     KamLAND2-Zen~\cite{shirai}                              & $^{136}$Xe  & $1000$  & $60$   & $[0,1.4]$   & $80$  & $97$             & $325$             & $2.1\cdot10^{\mbox{-}3}$ & $8.0\cdot10^{26}$ & $21$--$49$  & $15$   & $2$ & 2.9 \\
     nEXO~\cite{Mong}                                & $^{136}$Xe  & $4507$  & $25$   & $[\mbox{-}1.2,1.2]$ & $60$  & $85$             & $1741$            & $4.4\cdot10^{\mbox{-}4}$ & $4.1\cdot10^{27}$ & $9$--$22$   & $400$  & $1.2$ & 30 \\
     NEXT 100~\cite{Alvarez:2012flf,Martin-Albo:2015rhw} & $^{136}$Xe  & $91$    & $7.8$  & $[\mbox{-}1.3,2.4]$ & $88$  & $37$             & $26.5$            & $4.4\cdot10^{\mbox{-}2}$ & $5.3\cdot10^{25}$ & $82$--$189$ & n/a     & $1$ & 20 \\
     NEXT 1.5k~\cite{JJ} & $^{136}$Xe  & 1367    & 5.2  & $[\mbox{-}1.3,2.4]$ & $88$  & $37$             & 398            & $2.9\cdot10^{\mbox{-}3}$ & $7.9\cdot10^{26}$ & $21$--$49$ & n/a     & 1 & 300 \\
     PandaX-III 200~\cite{Chen:2016qcd}                & $^{136}$Xe  & $180$   & $31$   & $[\mbox{-}2,2]$     & $100$ & $35$             & $60.2$            & $4.2\cdot10^{\mbox{-}2}$ & $8.3\cdot10^{25}$ & $65$--$150$ & n/a     & n/a & n/a \\
     PandaX-III 1k~\cite{Chen:2016qcd}                & $^{136}$Xe  & $901$   & $10$   & $[\mbox{-}2,2]$     & $100$ & $35$             & $301$             & $1.4\cdot10^{\mbox{-}3}$ & $9.0\cdot10^{26}$ & $20$--$46$  & n/a     & n/a & n/a \\
     \hline
   \end{tabular}
   \label{tab:par}
\end{table*}
\begin{figure*}[tb]
   \centering
   \includegraphics[width=\textwidth]{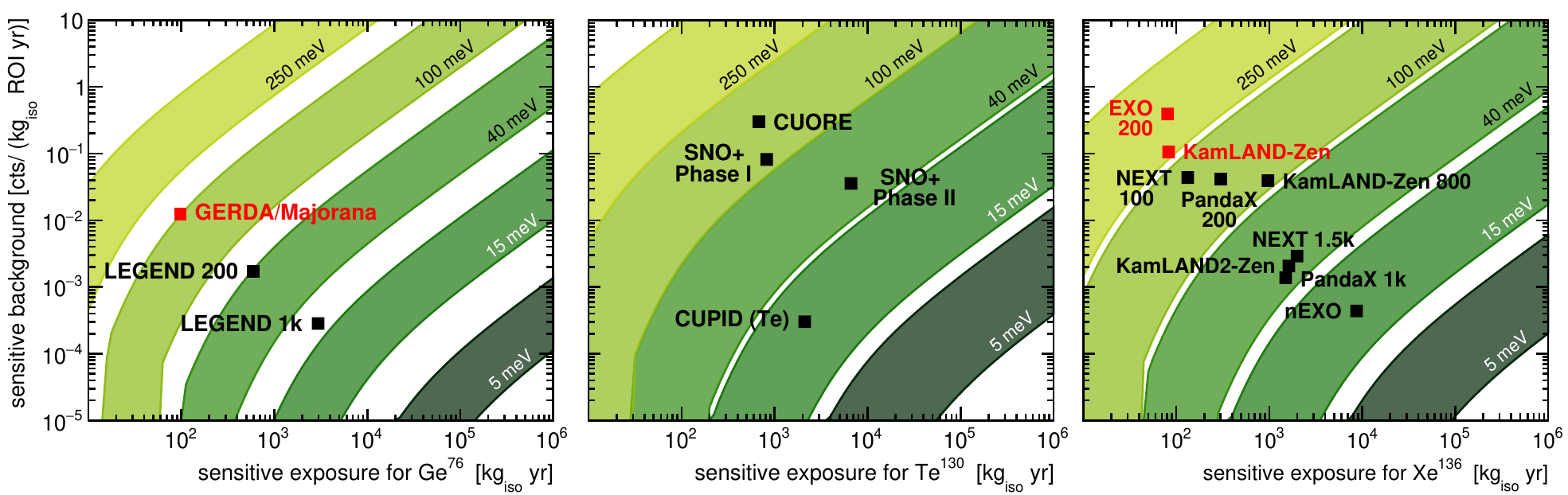}
   \caption{
     Discovery sensitivity for $^{76}$Ge, $^{130}$Te, and $^{136}$Xe as a function 
     of sensitive exposure and sensitive background. Contours in \mbb\ are represented
     as bands spanning the range of considered NME values. The experimental
     sensitivities of future or running experiments are marked after $5$
     years of live time. Past or current experiments with published background level and energy
     resolution (red marks) are shown according to the average performance in their
     latest data taking phase.}
   \label{fig:sensitivityBS}
\end{figure*}
With these considerations, the discovery sensitivity for each next-generation experiment
is computed using a heuristic counting analysis.
In cases where energy spectral fits and position non-uniformity enter non-trivially into the
sensitivity (as e.g. in SuperNEMO and nEXO), we tuned our parameters to match the collaboration's
stated sensitivity until agreement at the 10-20\% level was achieved. Again, our goal
is not to directly compare one experiment to another, but to interpolate the
sensitivity curves as a function of live time to allow a study of the discovery
probability of the ensemble of proposed experiments.
Further details of these computations and the input parameters are discussed in
\appendixname~\ref{sec:heuristics} and \ref{sec:parameters}. 

The sensitivity of an experiment to discover a signal is here defined as the
value of \hl\ or \mbb\ for which the experiment has a 50\% chance to measure a
signal with a significance of at least 3$\sigma$~\cite{Cowan:2010js}. 
\figurename~\ref{fig:sensitivityBS} plots the \mbb\ discovery sensitivity as a function of 
\senexp\ and \senbkg\ for three isotopes. Contours in \mbb\ are drawn as bands
representing the spread in NME for the given isotope.
The expected discovery sensitivity of each experiment after 5\,years 
of live time is marked in the plot and also included in \tablename~\ref{tab:par}. 
The \hl\ sensitivity after 10\,years of live time is about a factor $\sqrt{2}$ higher
for all experiments considered, although for the lowest background experiments
the improvement is as high as a factor of 1.6.
For each isotope, next-generation experiments are expected to reach discovery
sensitivity over the entire IO parameter space for at least some NME.

\section{Discovery probability}
\begin{figure*}[htb]
   \centering
   \includegraphics[width=\textwidth]{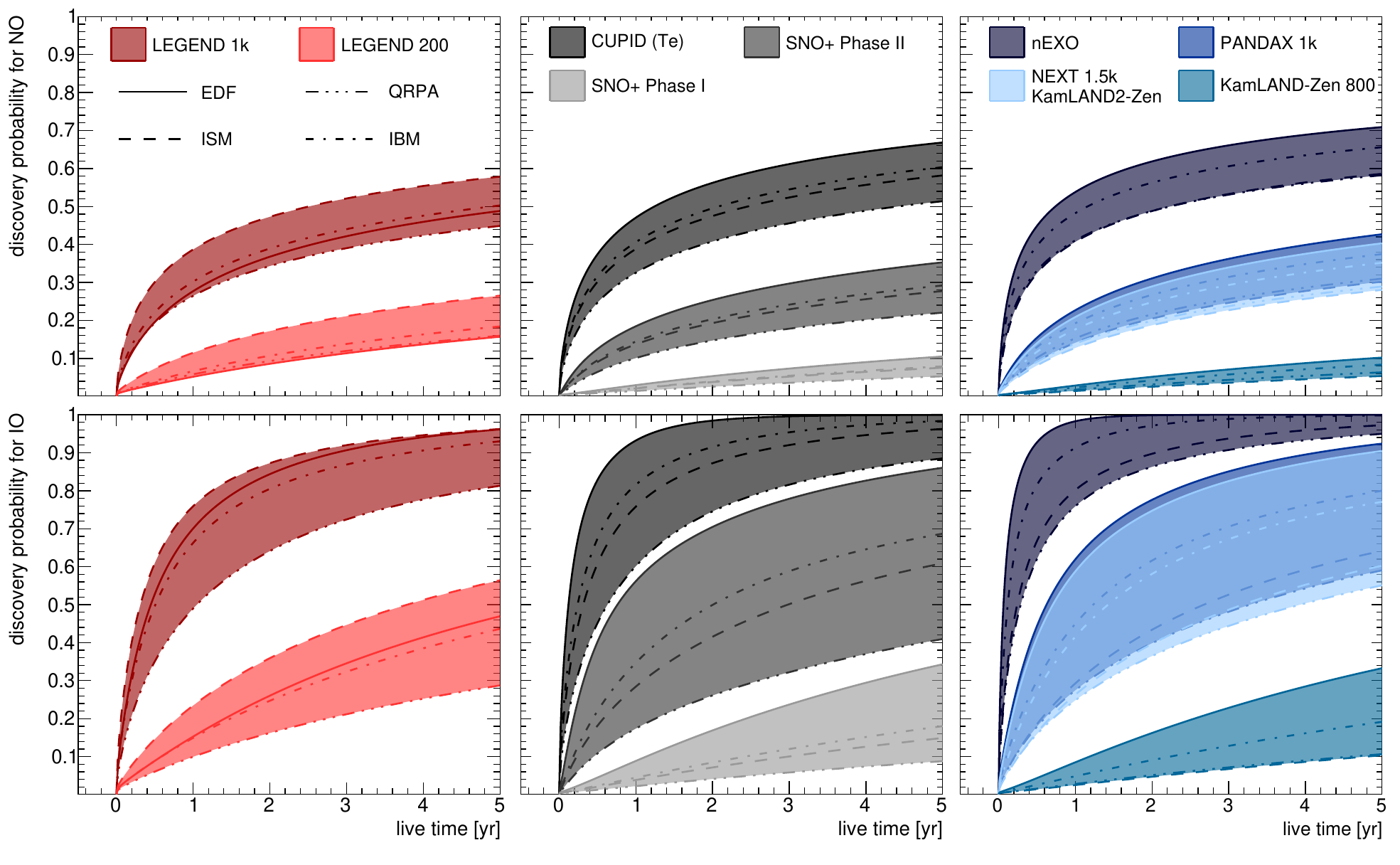}
   \caption{
     Discovery probability as a function of live time for a selection of
     next-generation experiments grouped according to the target isotope
     (from left to right: \isot{Ge}{76}, \isot{Te}{130}, \isot{Xe}{136}),
     assuming the absence of mechanisms that drive $m_l$ or \mbb\ to
     zero. The top
     panels show the discovery probability for NO, the bottom panels for IO.
     The variation of the NME among models is represented by the shaded regions. 
   }
   \label{fig:sensitivityTime}
\end{figure*}

The ultimate question that we want to address in this work is: what is the
probability of detecting a \nubb\ decay signal assuming that neutrinos are
truly Majorana particles?  
We define this Bayesian discovery probability as the odds of measuring 
a \nubb\ decay signal with a significance of at least 3$\sigma$. This is computed by
folding the discovery sensitivity with the probability distribution of \mbb\ output by the global fit. 
\figurename~\ref{fig:sensitivityTime} shows the evolution of the discovery
probability as a function of live time for a selection of next-generation
experiments, assuming the absence of mechanisms driving \mbb\ and $m_l$ to
zero. 
NME variations are visualized as bands.
The discovery probability for the most massive experiments 
exhibit a steep rise already in the first year or two of data taking. And while they begin
to flatten after 5 years, each experiment continues to gain discovery probability
out to 10 years.

For the case of Majorana neutrinos with an inverted mass ordering, 
next-generation experiments in each isotope are almost certain to observe a signal
in just 5 years of live time. Even several near-term experiments --
 SNO+~Phase\,I, KamLAND-Zen\,800, and LEGEND\,200 -- will have significant chances of 
discovery before the larger experiments come online.
Remarkably, several experiments also reach a discovery probability of
over 50\% in the case of NO. 
This strong possibility for discovery arises from the fact that even though the
NO parameter space extends down to exceedingly small \mbb, the amount of parameter 
space left at high \mbb\ comprises a significant fraction. 

One subtlety to note about the bands is that, while their width is driven by 
the NME variation, the relationship
between the discovery probability curves and the NME values is not monotonic. 
A change in NME model shifts the discovery sensitivity
to higher or lower \mbb\ for all isotopes. But the 
KamLAND-Zen and GERDA constraints result in 
parameter space opening up or being excluded at high \mbb\ 
depending on just the changes in the $^{136}$Xe and $^{76}$Ge NME values. 
This leads to both a shift and a subtle distortion of the \mbb\ probability
distribution. The ultimate change in the discovery probability is a 
non-trivial combination of these shifts and distortions.

We explored the impact on the discovery probabilities 
of adding also cosmological constraints to our global fit.
As expected based on the small deviations that these constraints generate in the \mbb\ distribution, 
we find that the discovery probability only degrades by $\sim$30\% for NO.
In the case of IO, the discovery probability of future experiments is so strong that
cosmological constraints are almost irrelevant.

We also explored the impact of $g_A$ quenching on the discovery probabilities.
Quenching degrades the sensitivity of future \nubb\ experiments, making parameter
space at low \mbb\ inaccessible. For a given half-life, the corresponding value of \mbb\ scales roughly like $g_A^{-2}$, so that
even just $\sim$30\% quenching can degrade an experiment's discovery sensitivity
by a factor-of-two. However, quenching also
relaxes the constraints imposed on \mbb\ by existing experiments, opening up
additional parameters space at high \mbb. As a result, the impact on discovery
potential isn't nearly as large as on the sensitivity. We find that a reduction of
$g_A$ by 30\% reduces the discovery power by $\sim$15\% ($\sim$25\%) for the most promising
future experiments in our reference analysis for IO (NO). 

When we include both 30\% quenching as well as
cosmological constraints, the region at high \mbb\ stays disfavored and the future
experiments simply lose reach. In this case we see this biggest suppression in discovery
power. However, even in this most pessimistic case the most promising experiments 
still have discovery power well above 50\% in the IO, and in the tens-of-percent range for NO.

The preceding discussion refers to our reference analysis.
To explore the case of extreme hierarchical neutrino masses, we 
repeat the analysis by fixing $\Sigma$ to its minimum allowed value. 
We find that for the IO the discovery probability is only marginally
impacted for the most promising next-generation experiments, decreasing by at most 10\%.
For the NO, as expected, the discovery probability drops to 2\% or lower for
all experiments and for all NME considered.

As a final note, we consider the impact of KATRIN~\cite{KATRIN}, which will perform a 
tritium-endpoint-defect-based kinematic measurement of the effective neutrino mass 
$m_\beta$ with 90\% CL limit-setting sensitivity of 200\,meV, 
and 5$\sigma$ discovery level of 350\,meV~\cite{Mertens:2016ihw}. 
An upper limit by KATRIN would have a marginal impact: 
including an $m_\beta$ 90\% upper limit at 200\,meV reduces discovery probabilities 
negligibly for the IO, and by less than 10\% for the NO. A discovery
by KATRIN, on the other hand, would be game-changing. Including a KATRIN
signal consistent with $m_\beta = 350$\,meV in our global analysis results in discovery
probabilities in excess of 99\% for both IO and NO, for all NME. 
In this case neutrino masses would be in the degenerate region, and 0$\nu\beta\beta$
decay experiments could not distinguish between NO and IO (cf.~\figurename~\ref{fig:correlations}).
Conversely, a non-observation of neutrinoless double-beta decay in that scenario would
reject the standard light, left-handed Majorana neutrino exchange mechanism
at high confidence level.

\section{Conclusions}
The probability distribution for the effective Majorana mass has
been extracted with a Bayesian global fit using all experimental information available
to-date and exploring various assumptions.
If the Majorana phases are not fixed by a flavor symmetry 
and the lightest mass eigenvalue is not driven to zero,
this distribution is found to peak at high values of \mbb\, not far from
existing limits. 
This puts much of the remaining parameter space within the reach of next-generation
experiments; it arises from the freedom of the Majorana phases and our requirement that the basis
choice yields a normalizable posterior distribution when scale-invariant priors are used.

The sensitivity of a suite of
next-generation \nubb\ decay experiments was estimated
with a heuristic counting analysis based on two parameters which fully determine the
performance of an experiment: the sensitive background and the sensitive
exposure. The sensitivity is finally combined with the probability distribution
of the effective Majorana mass to derive the discovery probability. 

The discovery probability is
found in general to be higher than previously considered for both mass orderings. 
For the inverted ordering, next-generation experiments will likely observe a signal
already during their first operational stages independently of the considered assumptions.
Even for the normal ordering, 
in the absence of neutrino mass mechanisms that drive the lightest state or the
effective Majorana mass to zero, the probability of discovering \nubb\  reaches
$\sim$50\% or higher in the most promising experiments.
These conclusions do not change qualitatively when cosmological constraints are imposed,
or when we allow for $g_A$ quenching. 

Our results indicate that even if
oscillation experiments or cosmological observations begin to strongly indicate a normal
ordering (as e.g.~in Ref. \cite{Simpson:2017qvj}),
next-generation \nubb\ decay experiments will still probe a relevant region of
the parameter space and give a valuable return on investment.  

~\\
{\bf Note added.} On the date of submission of our manuscript, a work by A.\,Caldwell
   and others~\cite{Caldwell:2017mqu} became public on the arXiv.
   They also perform a Bayesian global analysis of all data to extract a
   probability distribution for \mbb. Although their work has some similarities
   with ours, the most important difference is their use of the lightest
   mass eigenvalue ($m_{l}$) in the fit basis instead of $\Sigma$.
   When they use a flat prior for $m_{l}$, their results are in qualitative agreement with ours,
   including both discovery probabilities of future experiments as well as 
   the relative impact of cosmological constraints.
   However, when they use a log-flat prior (with cutoff set to $10^{-7}$\,eV),
   they find a degraded discovery probability, precisely as we discuss in
   Section~\ref{sec:globfit}.
   There are a few other key differences worth highlighting. Caldwell {\it et al.} extend their Bayesian
   analysis to include also priors on NME (primarily variations of QRPA) 
   as well as the mass orderings, which we treat independently.
   They do not consider quenching of the NME.  They also use a posterior odds threshold of 99\% 
   as the criterion for discovery, as opposed to our choice of a 3$\sigma$.   

\begin{acknowledgments}
The authors would like to thank
A. Caldwell,
S. R. Elliott,
G. Orebi Gann,
J.~J.~Gomez-Cadenas,
C. Pe\~na Garay,
G. Gratta,
Y. Mei,
J. Klein,
Yu. G. Kolomensky,
C. Licciardi,
L. Pandola,
D. Radford,
S. Sangiorgio,
B. Schwingenheuer,
S. Sch\"onert,
F. Vissani
and 
D. Waters
for valuable discussions and suggestions.
We are also grateful to the members of the \verb+nu-fit.org+ project 
(I.~Esteban, C.~Gonzalez~Garcia, M.~Maltoni, I.~Martinez~Soler, T.~Schwetz) 
to make their valuable results freely available to the community.
M.~A. acknowledges support by the Deutsche Forschungsgemeinschaft
(SFB1258).
\end{acknowledgments}

\appendix
\section{Global fit details and stability}\label{sec:fitDetails}
\begin{figure*}[tb]
   \centering
   \includegraphics[width=\textwidth]{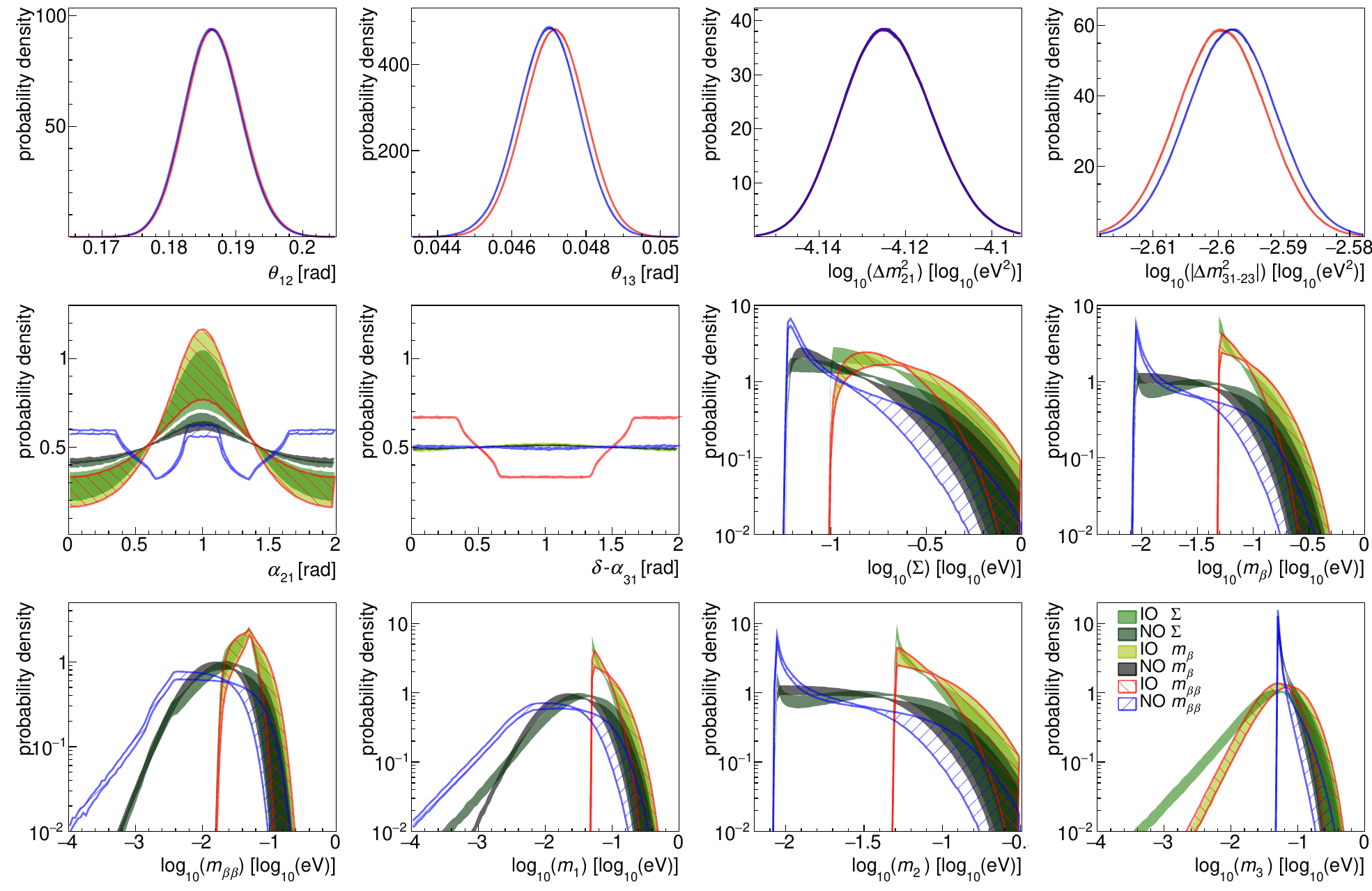}
   \caption{Marginalized posterior distributions for NO and IO.
      The band shows the deformation of the posterior distributions
      due to different assumptions on the NME.
      For each parameter three bands are displayed, corresponding to different
      parametrization of the basis in the fit. The lighter bands are obtained
      with the reference basis, the darker band are obtained for a basis in
      which $\Sigma$ is replaced by \mne, and the red and blue bands
      are obtained by replacing $\Sigma$ with \mbb.
   }
   \label{fig:allPost}
\end{figure*}

The marginalized posterior distributions for all the basis parameters used in
the global fit are shown in \figurename~\ref{fig:allPost}, along with the posterior
distributions for other physical parameters of interest. 
The bands show the deformation of the distributions due to the NME values.

\begin{figure*}[tb]
   \centering
   \includegraphics[width=\textwidth]{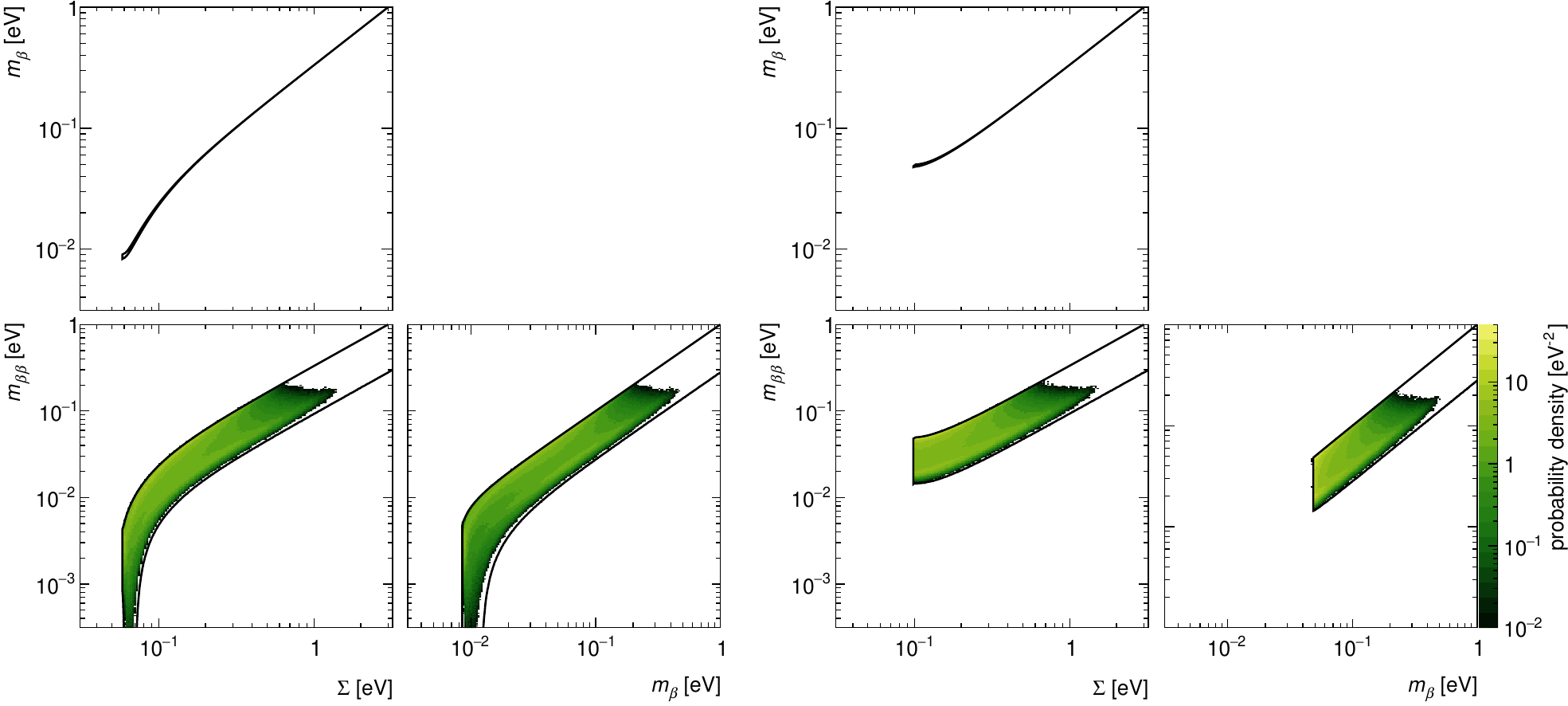}
   \caption{Marginalized posterior distributions for the mass observables
      assuming NO (left) and IO (right).The computation is performed assuming 
      QRPA NMEs and the absence of mechanisms
      that drive $m_l$ or \mbb\ to zero.
      The solid lines show the allowed parameter space assuming 3$\sigma$ intervals of the
      neutrino oscillation observables from nu-fit~\cite{Esteban:2016qun}.
      The probability density is normalized by the logarithms of the mass observables.}
   \label{fig:correlations}
\end{figure*}

The posterior distributions of the angles and mass splittings are Gaussian
and well defined. The shifts between IO and NO probability distributions
come from the results of oscillation experiments~\cite{Esteban:2016qun}.
The posterior distributions of the Majorana phases contain some information
as the current limits on \nubb\ decay force a partial cancellation
between the three terms on the RHS of equation~\eqref{eq:mbb}.  The posterior
distribution for $\alpha_{21}$ is more informative than for $(\delta-\alpha_{31})$
as the absolute value of the second term is larger than the third one.
The distributions of the parameters related to the mass eigenvalues are
considerably different for NO and IO as one would expect.
The posteriors of the mass eigenvalues and mass observables are constrained to a
finite range because of the relative volumes in the likelihood space.
For completeness, \figurename~\ref{fig:correlations} shows the 
correlations between the posterior distributions of the mass observables
$\Sigma$, \mbb\ and \mne, obtained assuming QRPA NMEs. 

There are three arbitrary elements in our analysis: the sets of data included in
the analysis, the priors, and the parametrization of the model (i.e.~the fit basis). 
The impact of the data set has been already discussed in the extreme case
in which Planck results are added to the analysis. This changes the posterior
distributions by about 10-20\%. 

The impact of the priors is in general weak as the data are strongly informative
for most of the parameters in the basis, with the exception of the Majorana phases and
$\Sigma$.  Using a flat prior for Majorana phases seems the only reasonable
choice: the parameter ranges  are well defined and no information is available
on them.  The prior used for $\Sigma$ is logarithmic to preserve scale
invariance.  An alternative choice would be a flat prior: this would favor larger
values of $\Sigma$ and inflate the discovery probabilities.

The parametrization of the model has potentially a huge impact on
the results. For this reason different parameterizations have been tested
and the impact on the posterior for \mbb\ was found to be in general marginal as
long as the basis covers all the degrees of freedom of the model and its parameters
are constrained by the data. 
For instance, \figurename~\ref{fig:allPost} shows the posterior distributions obtained
when $\Sigma$ is replaced by \mne\ in the fit basis (maintaining the logarithmic
prior). As occurs for $\Sigma$, lower \mne\ values are prohibited by
oscillation experiments and the parameter cannot vanish.
The posteriors are basically unchanged with the understandable
exception of $\Sigma$ and \mne.

Conversely, if a parameter of the basis is not sensitive to the data 
(e.g. $m_l$) its posterior probability coincides with the
prior and the resulting distribution cannot be normalized.
When $\Sigma$ is fixed to its minimum allowed value to study extreme hierarchical models
in which $m_l=0$, the posteriors become trivial:  all mass eigenvalues, $\Sigma$, and \mne\ 
are sharply peaked at their minimum allowed values, while posteriors for square mass differences and
angles are nearly indistinguishable from the results of the quasi-degenerate scenarios. 
Meanwhile, \mbb\ is pushed to lower values as shown in \figurename~\ref{fig:posteriors}.

Although there is no deep physical motivation to treat \mbb\ itself as a fundamental parameter,
for completeness we also performed the analysis substituting $\Sigma$ with \mbb\ in the parameter basis.
  For the IO, all posterior distributions except those of the Majorana phases coincide with the
  posteriors obtained with \mne\ in the basis.
  For the NO, \mbb\ could be vanishingly small, so the results depend in principle on the choice of the cutoff
  as for the case of the basis with $m_l$.
  However, with our flat priors for the Majorana phases, values of \mbb\ below $\sim$$10^{-3}$\,eV
  are strongly suppressed and the posterior of \mbb\ is not affected by the cutoff choice
  as long as it is $\lesssim10^{-5}$\,eV.
  The posterior for \mbb\ shows a preference for lower values with respect
  to what is obtained with our reference fit, which can be interpreted as a volume effect
  coming from the assignment of a log-flat prior on \mbb\ instead of on $\Sigma$.
  Additionally, the posteriors for the Majorana phases show different features with respect
  to what is obtained with the other bases. Again, these are the results of volume effects
  and indicate that the current knowledge does not allow us
  to make any clear statement on the Majorana phases.

\section{Heuristic counting analysis}\label{sec:heuristics}
In this work, the discovery sensitivity is defined to be the value of \hl\ or \mbb\ 
for which an experiment has a 50\% chance to measure a
signal above background with a significance of at least 3$\sigma$.
The computation is performed for \hl\ and the result converted to a range of \mbb\
values by using equation~\eqref{eq:t12tombb} with different NME values.
Given an expectation for the background counts in the \senroi\ of $B= \senbkg
\senexp$,
the sensitivity for \hl\ is given by:
\begin{equation}
\hl\ = \ln2 \frac{N_A \senexp}{m_a S_{3\sigma}(B)} ,
\label{eq:discsens}
\end{equation}
where $S_{3\sigma}(B)$ denotes the Poisson signal expectation
at which 50\% of the
measurements in an ensemble of identical experiments would report a 3$\sigma$
positive fluctuation above $B$. 
If $B$ is large then $S_{3\sigma}(B) \propto \sqrt{B}$, while if
$B(t) \ll 1$ then $S_{3\sigma}(B)$ is a constant.
To transition smoothly between
these two regimes, we find the number of counts $C_{3\sigma}$ such that the cumulative
Poisson distribution with mean $B$ satisfies CDF$_{Poisson}(C_{3\sigma}|B) =
3\sigma$, and then obtain $S_{3\sigma}$ by solving  
$\overline{\textrm{CDF}}_{Poisson}(C_{3\sigma}|S_{3\sigma}+B) = 50\%$, as suggested in \cite{Punzi:2003bu}
(where $\overline{\textrm{CDF}}$ refers to the complementary CDF).
While $C_{3\sigma}$ should strictly be integer-valued, restricting it as such
would result in discrete jumps in the discovery sensitivity as $B$
increases.
To smooth over these jumps we extend CDF$_{Poisson}$
to a continuous distribution in $C$ using its definition via the normalized
upper incomplete gamma function:
\begin{equation}
\textrm{CDF}_{Poisson}(C|\mu) = \frac{\Gamma(C+1, \mu)}{\Gamma(C+1)}.
\label{eq:smoothCDF}
\end{equation}
Using equation~\eqref{eq:smoothCDF}, $S_{3\sigma}$ varies smoothly and
monotonically with $B$ for values greater than $-\ln\left[erf(3/\sqrt{2})\right] = 0.0027$~counts.
Below this value of $B$, the observation of a single count represents a 3$\sigma$ discovery,
marking this as the level at which an experiment becomes effectively ``background-free''
under this metric. In this regime, $S_{3\sigma}$ takes the constant value $\ln 2$.

\begin{figure}[tb]
   \centering
   \includegraphics[width=\columnwidth]{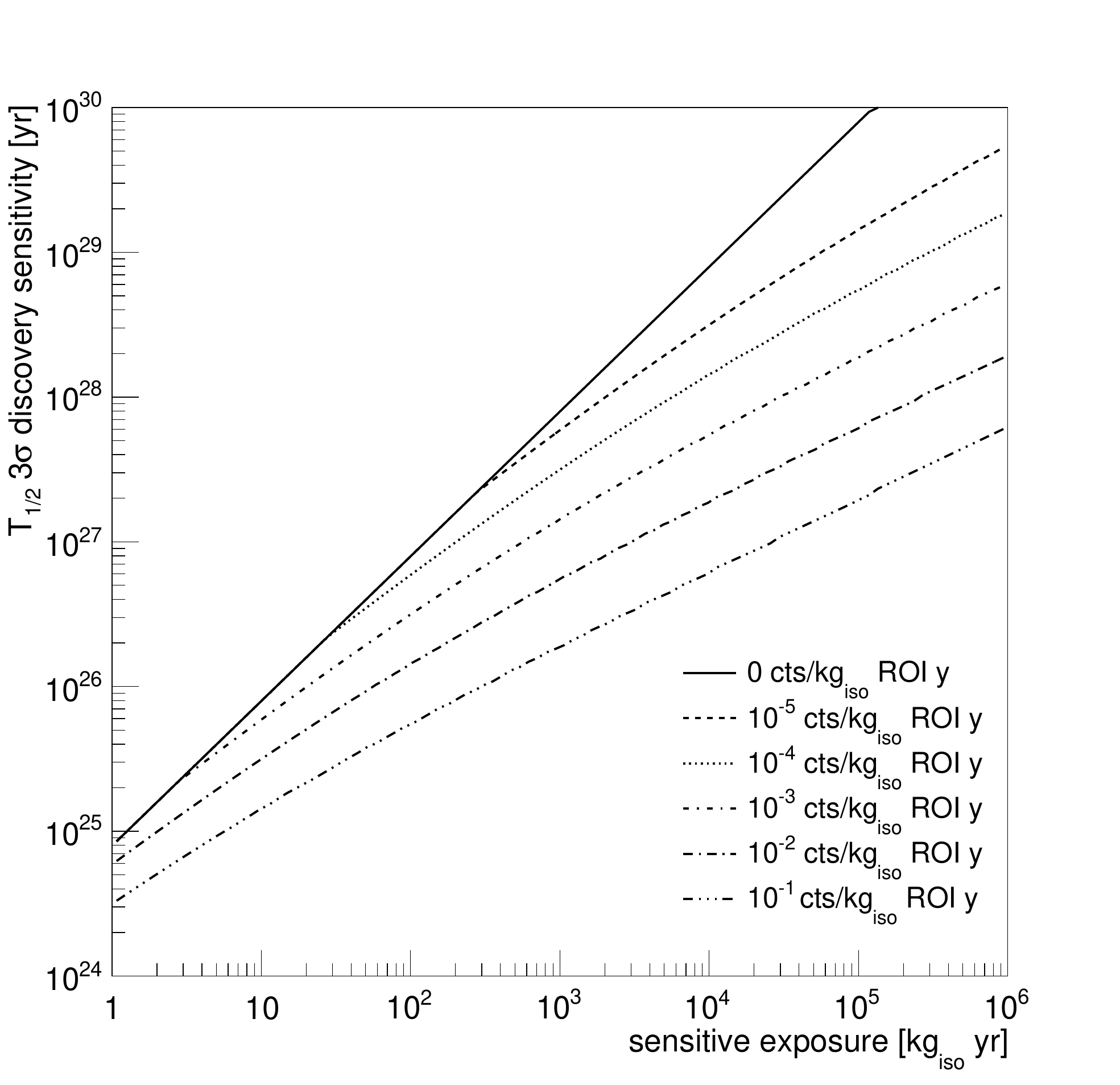}
   \caption{$^{76}$Ge \hl\ discovery sensitivity as a function of sensitive exposure for
   a selection of sensitive background levels.}
   \label{fig:discsens}
\end{figure}

Using equations~\eqref{eq:discsens} and~\eqref{eq:smoothCDF},
the \hl\ sensitivity for \isot{Ge}{76} as a function of $\senexp$ and $\senbkg$ is shown in
\figurename~\ref{fig:discsens}. Values for other isotopes can be obtained by
dividing by the ratio of their molar mass to that of $^{76}$Ge.
Discovery sensitivity increases linearly with exposure
until the experiment exceeds the background-free threshold of 0.0027 counts.
For a given exposure, the sensitivity degrades rapidly with background level.

In a similar manner, the discovery probability is defined to be the probability that
an experiment will measure a 3$\sigma$ positive fluctuation above $B$, given the
probability distribution function $dP/dm_{\beta\beta}$ for \mbb\ (i.e.~\figurename~\ref{fig:posteriors}). 
Explicitly, the discovery probability (DP) is computed as
\begin{equation}
\textrm{DP} = \int_0^\infty \frac{dP}{dm_{\beta\beta}} \,  \overline{\textrm{CDF}}_{Poisson}(C_{3\sigma}|S(m_{\beta\beta})+B) \, dm_{\beta\beta},
\end{equation}
where $S(m_{\beta\beta})$ is the expected signal counts in the experiment for a given value of $m_{\beta\beta}$.

For high resolution experiments with flat background spectra in the vicinity of
the $Q$ value, we performed an optimization of the \senroi\ width by 
maximizing the figure-of-merit
\begin{equation}
\textrm{F.O.M.} = \frac{erf\left(n/\sqrt{2}\right)}{S_{3\sigma}(bn)}
\end{equation}
where $n$ is the \senroi\ half-width in units of the energy resolution ($\sigma$), 
and $b$ is the background counts per unit $\sigma$ at 5 years of live time.
Since $S_{3\sigma}(bn) \propto \sqrt{bn}$ for large values of $b$, in this
regime the F.O.M.~is maximal for the value of $n$ that solves the transcendental
equation $n e^{-n^2/2} = erf\left(n/\sqrt{2}\right) \sqrt{\pi} / 4$.
This gives an optimal \senroi\ width of 2.8$\sigma$ for background-dominated experiments,
with a corresponding signal efficiency of 84\%.
At lower background the sensitivity improves with a wider \senroi. 
In the background-free regime, the F.O.M.~is optimized when the \senroi\ width
is expanded until the region contains 0.0027 count.
Above this region, the F.O.M.~was maximized numerically, making
use of equation~\eqref{eq:smoothCDF}. The deviations from the asymptotic
value of 2.8 were plotted on a log-log scale and were found to be
well-approximated by a 2$^{nd}$-order polynomial. This gives the following
expression for the optimum \senroi\, accurate to $<$1\%:
\begin{equation}
\textrm{ROI}_{\textrm{opt}} = 2.8 + 10^{a_0 + a_1 \log_{10}b + a_2 \log_{10} 2b}
\end{equation}
where the parameter values are $a_0 = -0.40$, $a_1 = -0.29$, and $a_2 = -0.039$.

Our treatment ignores uncertainty in the background rate as well as
systematic uncertainties. Backgrounds are typically well-constrained in
\nubb\ experiments using sidebands in energy and, for some detectors,
position. Similarly, systematic uncertainties are typically well below 10\%.
This makes these sources of uncertainty subdominant to the large fluctuations
that drive low-count-rate Poisson statistics.

\section{Experimental parameters}\label{sec:parameters}
This appendix discusses the experiments and parameters listed in
\tablename~\ref{tab:par}.
The parameter values are taken from official publications and presentations of
each collaboration. If not available, the values are assumed to be
the same of predecessor or similar experiments (e.g. the instrumental efficiency
is usually not given prior to the construction and operation of an experiment).
Our heuristic counting analysis is used to derive the sensitivity of each experiment for
both a limit setting and a signal discovery analysis~\cite{Cowan:2011an}.
The collaborations typically
quote only the former, but this is enough to cross-check -- and possibly tune --
the sensitive background and exposure used for this work.
Given the values in \tablename~\ref{tab:par}, our calculation reproduces the
official sensitivities quoted by each experiment with 10-20\% accuracy.

LEGEND~\cite{bernhard,stefan}
is the successor of GERDA and {\sc Majorana}~\cite{Ackermann:2012xja,Abgrall:2013rze}.
The project consists of two stages: LEGEND 200 and LEGEND 1k. 
In the first phase, $200$\,kg of germanium detectors enriched at $87\%$ in $^{76}$Ge
will be operated in the existing GERDA infrastructure. 
The background level measured in GERDA Phase\,II is
\senbkg$=$1.2$\cdot10^{-2}$\,\ctssen\ in average and 
5.1$\cdot10^{-3}$\,\ctssen\ when only the new generation BEGe-type
detectors are considered~\cite{stefan}.
Compared to the results obtained with BEGe detectors, 
a further reduction of a factor $\sim$3 is expected in LEGEND 200.
For LEGEND 1k, a new infrastructure able to host $1$\,ton of target mass and a further
6-fold background reduction are conceived.
We assume the same resolution achieved by the running experiments
($\sim$3\,keV full width at half maximum, FWHM),
and use a \senroi\ of ($Q$-value $\pm2\sigma$).
Enrichment, active volume, containment and instrumental
efficiency are taken from Ref.~\cite{Agostini:2017iyd}.
Our calculation agrees with the sensitivity projections of the
collaboration~\cite{bernhard,stefan} when the same \senroi\ is used.

SuperNEMO~\cite{Povinec:2017trz,Arnold:2010tu}
is an upgrade of the NEMO-3 experiment. 
This is the only experiment considered here in which the \nubb\ decay isotope is
separate from the detector. SuperNEMO will consist of $20$ identical tracking chambers, each 
containing $\sim$5\,kg of $^{82}$Se embedded in Mylar foils.
SuperNEMO can measure independently the energy
and direction of the two electrons emitted by \nubb\ decays, 
and distinguish different decay channels~\cite{Arnold:2010tu}.
The electrons do not release all their energy
in the chamber: the expected \nubb\ decay signature for $^{82}$Se is thus a Gaussian peak
at $\sim$2830\,keV, about $170$\,keV below the $^{82}$Se $Q$-value~\cite{Arnold:2010tu}.
The product of containment and instrumental efficiency
for $^{82}$Se \nubb\ decay events is quoted to be $28.2\%$ in
Ref.~\cite{Arnold:2010tu}.
However, the \nubb\ decay peak
will be on the tail of the \nunubb\ decay spectrum
(see \figurename~5 of Ref.~\cite{Arnold:2010tu}).
We therefore extracted the expected total efficiency and total number of background counts
for different energy ranges, and use the ones providing the best sensitivity,
i.e.~[$2800,3100$]\,keV. 
The corresponding total efficiency, which also includes the fraction of \nubb\ decay
events falling within the \senroi, is taken to be $16.5\%$.
With such parameters, we accurately reproduce the official sensitivity~\cite{Arnold:2010tu}.
SuperNEMO expects to improve their energy resolution by
a factor of $2$ and the background level by a factor of $\sim$50 with respect to NEMO-3~\cite{Povinec:2017trz,Arnold:2010tu}.

CUPID~\cite{Wang:2015raa,Wang:2015taa,Artusa:2014wnl}
is an upgrade of the CUORE experiment~\cite{CUORE-NIM2004,Artusa:2014lgv}.
In CUORE, $\sim$1000 TeO$_2$ crystals with natural isotopic composition are operated as calorimeters (bolometers)
at a base temperature of $\sim$10\,mK.
CUPID plans to exploit the CUORE cryogenic infrastructure,
and increase the sensitivity to \nubb\ decay using enriched crystals with $\alpha/\beta$ discrimination capabilities.
Several crystals with different double-$\beta$ decaying isotopes are under investigation,
including TeO$_2$, ZnMoO$_4$, ZnSe and CdWO$_4$.
We quote results only for  TeO$_2$ and ZnSe, which we found to yield the lowest background and the highest
sensitivity.
Both CUORE and CUPID aim at an energy resolution of $\sim$0.2\% (FWHM), which
has been proven on a large array of TeO$_2$ crystals in CUORE-0~\cite{Alduino:2016zrl}.
In CUORE, a background level reduction of a factor $\sim$6 with respect to CUORE-0 is expected
thanks to improved shielding and a careful selection of all materials~\cite{Alduino:2017qet}.
A further reduction in background level by a factor $\sim$500 is conceived for CUPID with TeO$_2$:
this can be achieved thanks to the readout of Cherenkov light induced by electrons in TeO$_2$,
or of the scintillation light in the other crystals mentioned above.
The optimal \senroi's for CUORE and CUPID are ($Q$-value$\pm1.4\sigma$) and ($Q$-value$\pm2\sigma$), respectively.
For both experiments we used an instrumental efficiency of $92\%$ as in its predecessor CUORE-0~\cite{Alduino:2016zrl}.
The exclusion sensitivity we obtained differs by $\lesssim$10\%
from the official values~\cite{Alessandria:2011rc,Wang:2015raa}.

SNO+ is an ongoing upgrade of SNO.
It is a multi-purpose neutrino experiment, with \nubb\ decay search
as one of its main physics goals~\cite{Lozza:2016rwo,singh}.
An acrylic sphere with about $800$\,tons of liquid scintillator,
loaded with tellurium, will be inserted in water.
A multi-staged approach is foreseen. 
In SNO+ Phase\,I, $\sim$1.3\,tons of $^{130}$Te
are used and an energy resolution of 7.5\% FWHM is expected.
The goal of SNO+ Phase\,II~\cite{KleinSNO2}
is to increase the $^{130}$Te mass to $\sim$8\,tons
and improve the energy resolution to 5.3\%.
This is achievable thanks to an improvement
of the light yield to $800$\,pe/MEV~\cite{KleinPrivate}.
We assumed a containment efficiency of $100\%$ 
and an instrumental efficiency of $\sim$97\% as for KamLAND-Zen.
Using an asymmetric \senroi\ of
($Q$-value $^{+1.5}_{-0.5}\sigma$)~\cite{Lozza:2016rwo,singh,KleinSNO2},
we reproduce the official limit-setting sensitivity\cite{singh} with a few
percent accuracy.

KamLAND-Zen is a KamLAND upgrade
tailored to the search of \nubb\ decay: a nylon balloon is inserted in the active detector volume
and filled with liquid scintillator loaded with enriched xenon.
After two successful data taking phases~\cite{Gando:2012zm,KamLAND-Zen:2016pfg},
the KamLAND-Zen collaboration is currently preparing two additional phases
called KamLAND-Zen 800 and KamLAND2-Zen
in which $750$\,kg and $1$\,ton of $^{136}$Xe will be deployed, respectively.
A major upgrade of the experiment is conceived for KamLAND2-Zen to
improve the energy resolution at the $^{136}$Xe $Q$-values from $4.6\%$ to $2\%$ ($\sigma$) and to reduce the
background by an order of magnitude. The upgrade includes the installation of new light
concentrators and PMTs with higher quantum efficiency~\cite{shirai} as well as
purer liquid scintillator.
In our study we used the same instrumental efficiency as reported in
Ref.~\cite{KamLAND-Zen:2016pfg}.
The optimal \senroi\ is asymmetric covering only the upper half of the
expected \nubb\ decay peak to avoid the background due to the \nunubb\ decay spectrum tail.
Our calculations reproduce the sensitivities presented 
in~\cite{Gando:2012zm,KamLAND-Zen:2016pfg,shirai} within 20\%.
The background measured in KamLAND-Zen phase 2 is
\senbkg$=$1.1$\cdot10^{-1}$\,\ctssen\ in average, and
5.9$\cdot10^{-2}$\,\ctssen\ when only the second part of
the data taking is considered (period 2).
Compared to this last result, a further reduction of a factor 1.5 ($\sim$15) is expected for KamLAND-Zen
800 (KamLAND2-Zen).

nEXO~\cite{Pocar:2015mrz}
is an upgrade of the EXO-200~\cite{Albert:2014awa} experiment.
The detector is a liquid Time Projection Chamber (TPC) filled with $5$~tons of xenon enriched at $90\%$ in $^{136}$Xe.
One of the main background contributions expected in nEXO is due to radioactive
isotopes in the TPC materials. Because of the self-shielding of the Xe material,
the rate of background events decreases exponentially moving toward the center.
The collaboration plans to perform 
an analysis of the full detector
volume, using the outer part to constrain the external background contribution.
Our counting analysis cannot take care of this and we are forced to tune the sensitive
background and exposure.
Given a fiducial volume of 3\,tons of Xe, a \senroi\ of ($Q$-value$\pm1.2\sigma$) and an
average background level of $\sim$4$\cdot10^{-6}$\,cts/keV/kg$_{iso}$/yr
(that is $\sim$20\% of the reference value~\cite{Mong:2016sza,yang}) we obtain a
discovery sensitivity 15-20\% lower than the collaboration's estimate. This
is however sufficient for our analysis.
The instrumental efficiency is taken for EXO-200~\cite{Albert:2014awa}.
In nEXO, the energy resolution is expected to be improved by a factor of $1.2$, and the background level
reduced by about a factor $400$ with respect to EXO-200, due primarily to better self-shielding 
and more efficient background identification in the larger experiment.

NEXT~\cite{Alvarez:2012flf}
aims at searching for \nubb\ decay using a high-pressure Xe-gas TPC,
which combines tracking capabilities with a low background typical of experiments
with a single element in the active volume.
The expected presence
of the $^{214}$Bi gamma line at 2447\,keV in vicinity of the \nubb\ decay $Q$-value at 2458\,keV,
requires the use of an asymmetric \senroi~\cite{Martin-Albo:2015rhw}.
A single TPC with 100\,kg of Xe ($90\%$ $^{136}$Xe) and a resolution of
0.75\%\,FWHM~\cite{Martin-Albo:2015rhw} will be used in the next phase of the
project (NEXT 100). In a later stage, the collaboration plans to operate an array of 3
TPCs, each with a total Xe mass of 500\,kg, a background level lower by a factor $\sim$10
with respect to NEXT 100 and an improved energy resolution of
0.5\%\,FWHM~\cite{JJ} (NEXT 1.5k)~\cite{JJ}.
The total efficiency is taken from~\cite{Martin-Albo:2015rhw}:
the value reported in \tablename~\ref{tab:par} does not contain the fiducial volume fraction ($88\%$)
and the fraction of events in the \senroi\ ($90\%$).
We compared the NEXT 100 exclusion sensitivity obtained with our approach
with that given in~\cite{Martin-Albo:2015rhw}, and find that the two values
agree within $\sim$10\%.

Another experiment using the same technique of NEXT is PandaX.
After two phases dedicated to dark matter searches,
a \nubb\ decay search program -- denoted PandaX-III -- is planned~\cite{Chen:2016qcd}.
The TPC of PandaX-III will be about twice as big as that of NEXT,
but will have an energy resolution of about $3\%$\,FWHM~\cite{Chen:2016qcd}.
As for NEXT, one of the major expected backgrounds is $^{214}$Bi.
Consequently, an asymmetric \senroi\ would yield a higher sensitivity,
but for consistency with Ref.~\cite{Chen:2016qcd} we used an \senroi\ of ($Q$-value$\pm2\sigma$).
We could not find information regarding the size of the fiducial volume,
and we assume it to be $100\%$.
The total efficiency is about $35\%$~\cite{Chen:2016qcd}.
In a second stage, the PandaX-III collaboration plans to construct  four additional TPCs with  
energy resolution improved to $1\%$\,FWHM and a background level reduced by one order of magnitude.
Our evaluation of the exclusion sensitivity
agrees at the $\sim$10\% level with the official value~\cite{Chen:2016qcd}.

\bibliography{saving-nubb-bib}

\end{document}